# On the super-fluid property of the relativistic physical vacuum medium and the inertial motion of particles

Shun-Jin Wang

*Department of Physics, Sichuan University, Chengdu 610064, PR China*

Email address of the author: sjwang@home.swjtu.edu.cn.

**Abstract:** The similarity between the energy spectra of relativistic particles and that of quasi-particles in super-conductivity BCS theory makes us conjecture that the relativistic physical vacuum medium as the ground state of the background field is a super fluid medium, and the rest mass of a relativistic particle is like the energy gap of a quasi-particle. This conjecture is strongly supported by the results of our following investigation: a particle moving through the vacuum medium at a speed less than the speed of light in vacuum, though interacting with the vacuum medium, never feels friction force and thus undergoes a frictionless and inertial motion. The profound and intrinsic relationship between the super fluid property of the relativistic physical vacuum medium and the energy-momentum conservation law as well as the relativistic energy-momentum dispersion relation or the principle of relativity, can be established.

**Pacs numbers: 03.65.-w, 03.65.Ca, 03.65.Pm, 11.30.Er**

The property of the physical vacuum of our universe is a central issue in modern particle physics and cosmology. In this paper we shall show that the relativistic physical vacuum medium as a ubiquitous back ground field is a super fluid medium.

Our ground of argument is based on the extension of Landau's talent argument on the super-fluid problem[1,2]. Our starting points of argument are as follows: (1) the particle has a relativistic energy spectrum; (2) the energy-momentum conservation law is observed in the physical process. The conclusion is that the motion of the particle in the vacuum medium is frictionless and inertial, and the relativistic physical vacuum medium is a super-fluid medium.

Consider a particle with the rest mass $M$ and moving at the velocity $\vec{V}$. Its energy and momentum read,

$$E = Mc^2/\sqrt{1-\beta^2}, \qquad \vec{P} = M\vec{V}/\sqrt{1-\beta^2} \qquad (1)$$

where $\beta = |V|/c$. Assume that as the particle interacts with the background vacuum medium, a particle is created from the vacuum medium with the momentum $\vec{p}$ and the energy $\varepsilon(\vec{p})$.

According to Landau's argument, the dissipation is resulted microscopically due to the creation of the new particles from the medium, both the particle creation process and the vacuum medium are thus quantum in nature. After momentum and energy loss, the particle $M$ has the velocity $\vec{V}_1$, momentum $\vec{P}_1$, and energy $E_1$, which are related by the relativistic relations,

$$E_1 = Mc^2/\sqrt{1-\beta_1^2}, \quad \vec{P}_1 = M\vec{V}/\sqrt{1-\beta_1^2} \qquad (2)$$

where $\beta_1 = |V_1|/c$. According to the energy-momentum conservation, one has

$$Mc^2/\sqrt{1-\beta^2} = Mc^2/\sqrt{1-\beta^2} + \varepsilon(\vec{p}) \qquad (3)$$

$$M\vec{V}/\sqrt{1-\beta^2} = M\vec{V}/\sqrt{1-\beta_1^2} + \vec{p} \qquad (4)$$

From eq.(4), one obtains,

$$\frac{1}{1-\beta_1^2} = 1 + [\frac{M\vec{V}}{\sqrt{1-\beta^2}} - \vec{p}]^2 / M^2 c^2$$
$$= \frac{1}{1-\beta^2} + [(\frac{p}{Mc})^2 - 2(\frac{p}{Mc})\frac{\beta \cos\theta}{\sqrt{1-\beta^2}}] \qquad (5)$$

where $\theta$ is the angle between $\vec{V}$ and $\vec{p}$ such that $\vec{V} \cdot \vec{p} = VP\cos\theta$. Inserting eqs.(5) into eq.(3), we have

$$\frac{2Mc^2(VP\cos\theta - \varepsilon(\vec{p}))}{\sqrt{1-\beta^2}} + (\varepsilon(\vec{p})^2 - c^2 p^2) = 0. \qquad (6)$$

Equation (6) expresses the physical constraint on the physical process from the energy-momentum conservation: only the process which obeys eq.(6) is physically realizable.

Now we continue our discussion in two cases:

(1) $M$ is the rest mass of a macroscopic particle, while $\{\varepsilon(\vec{p}), \vec{p}\}$ are the energy and momentum of a microscopic particle. As was done by Landau, the second term of the left hand side of (6) can be neglected, and this leads to

$$Vp\cos\theta = \varepsilon(\vec{p}), \quad V \geq \varepsilon(\vec{p})/p \qquad (7)$$

For the relativistic particle, we have $\varepsilon(\vec{p}) = pc$ or $\varepsilon(\vec{p}) = \sqrt{m^2 c^4 + c^2 p^2}$ which leads to

$$[\varepsilon(\vec{p})/p]_{min} = c \qquad (8)$$

From eqs.(7) and (8), we obtain

$$V \geq c \qquad (9)$$

which is impossible for the relativistic dynamics. Therefore, a macroscopic particle moving at the speed less than $c$ in the vacuum medium can never excite microscopic particles from the vacuum medium, thus it will undergo a frictionless and inertial motion without any energy and momentum loss.

(2) Both $M$ and $m$ are the rest masses of microscopic particles, and

$$\varepsilon(\vec{p}) = \sqrt{m^2 c^4 + c^2 p^2}, m \geq 0 \qquad (10)$$

From the energy conservation eq.(3), one has

$$\frac{1}{1-\beta_1^2} = \frac{1}{1-\beta^2} + (\frac{m}{M})^2 + (\frac{p}{Mc})^2 + \frac{2}{\sqrt{1-\beta^2}}\sqrt{(\frac{m}{M})^2 + (\frac{p}{Mc})^2} \qquad (11)$$

and from the momentum conservation eq.(4), one has eq.(5). Comparing eqs.(11) and (5), we obtain

$$(\frac{m}{M})^2 = -\frac{2}{\sqrt{1-\beta^2}}[\sqrt{(\frac{m}{M})^2 + (\frac{p}{Mc})^2} + (\frac{p}{Mc})\beta\cos\theta] < 0 \qquad (12)$$

which indicates that under the energy-momentum conservation, equation (12) can not yield a real particle mass $m$. Therefore the physical process is forbidden by the energy-momentum conservation.

From the above investigation, we conclude that a particle with the rest mass $M$ and relativistic energy spectrum, moving in the vacuum medium at the speed less than $c$, can not excite particles from the vacuum medium and will not feel friction force, thus it will undergo a frictionless and inertial motion. This is a direct consequence of the energy-momentum conservation and the relativistic energy spectra of the particle. The result further implies that the relativistic physical vacuum medium under the condition of energy-momentum conservation is a super-fluid (frictionless) medium for the particles with relativistic energy spectra and moving through it at a speed less than $c$. On the contrary, if the particle would move at the speed larger than $c$, it would lose energy to the vacuum medium as the Cherenkov-type radiation. That is why no particle can keep its speed larger than $c$.

The above proof has been done on a special reference frame where the vacuum medium is rest and the particle is moving respect to the vacuum medium. In this proof, the vacuum medium is assumed to be a special material and the particle is moving through this material. The friction force (if it exists) of the vacuum medium on the particle comes from the interaction between the particle and the vacuum medium. This interaction can make the fast moving particle excite new particles from the vacuum medium, lose its energy and slow down. This might be a friction and dissipation process. This picture is physically intuitive and directly related to the Landau's picture. In the proof, only the energy-momentum conservation law and the relativistic energy-momentum dispersion relation are used. The principle of relativity and the corresponding Lorentz ( Poincare ) invariance are not referred directly.

However, the proof can be made in an alternative way which is simple but indirect , based on the Einstein's special principle of relativity and Lorentz transformation besides the energy-momentum conservation. Suppose that the particle $M$ is rest with respect to the vacuum medium. It is evident that the particle $M$ can not create a new particle from the vacuum medium and simultaneously keep its same rest mass $M$ since the new particle costs extra energy and the process thus violates energy conservation. Boost the particle $M$ to obtain a constant velocity $\vec{V}$ with respect to the vacuum medium by the Lorentz transformation. According to the Einstein's special principle of relativity, the moving particle should obey the same physical law as it is rest so that the moving particle can not excite a new particle either from the vacuum medium because the same energy conservation law is still valid for the moving particle. This in turn implies that the moving particle $M$ should undergo an inertial motion, namely there is no friction force exerting on the particle even it is moving at the speed $V$ less than $c$ with respect to the vacuum medium. The difference of the proof here is that its starting points are the Einstein's special principle of relativity, the Lorentz transformation, and the energy conservation, which lead to the same conclusion that the vacuum medium is a super-fluid medium.

It is clear that these two proofs are equivalent to each other, since the Einstein's special principle of relativity and the Lorentz transformation lead to the relativistic energy-momentum dispersion relation which is one of the two bases in the first proof.

It is very important to note that the Galileo-Newton vacuum medium, on the contrary, does not have the above super fluid property. This is because that the Galileo-Newton vacuum medium does not yield a particle energy spectrum with energy gap. The statement can be proved as follows. Suppose the above process takes place in the Galileo-Newton vacuum medium which yields the following energy spectra with the energy-momentum relation

$$E = \frac{P^2}{2M}, \ \vec{P} = M\vec{V}; \quad E_1 = \frac{P_1^2}{2M}, \ \vec{P}_1 = M\vec{V}_1; \quad \varepsilon(\vec{p}) = \frac{p^2}{2m}, \ \vec{p} = m\vec{v} \qquad (13)$$

clearly showing that there is no energy gap. The energy-momentum conservation of the above process reads:

$$\frac{1}{2}MV^2 = \frac{1}{2}MV_1^2 + \frac{p^2}{2m}, \qquad M\vec{V} = M\vec{V}_1 + \vec{p} \qquad (14)$$

From eq.(14), we have the following equation to determine $V_1$

$$(\frac{M}{m}+1)V_1^2 - (2\frac{M}{m}V\cos\theta)V_1 + (\frac{M}{m}-1)V^2 = 0 \qquad (15)$$

where $\theta$ is the angle between $\vec{V}$ and $\vec{V}_1$ such that $\vec{V}\cdot\vec{V}_1 = VV_1\cos\theta$. The solution of eq. (15) is

$$V_1 = V[\frac{M}{m}\cos\theta \pm \sqrt{1-(1-\cos^2\theta)(\frac{M^2}{m})}]/(\frac{M}{m}+1) \qquad (16)$$

with the real solution condition $(1-\cos^2\theta)(\frac{M^2}{m}) \leq 1$. It is clear that $\vec{V}_1(V_1,\theta)$ depends on $(M,m)$ and the dissipative process is characterized by the quantities $\vec{V}_1(V_1,\theta)$ and $m$. Consider a simple possibility where $\theta = 0$. Then the solutions of eq. (16) are (i) $V_1 = V, p = 0$ without dissipation since $\varepsilon(\vec{p}) = 0$; and (ii) $V_1 = V(M/m-1)/(M/m+1)$, with the dissipation energy $\varepsilon(\vec{p}) = [2(MV)^2/m]/(M/m+1)^2$. In general, there are many other possibilities where $V_1 < V$ and $\varepsilon(\vec{p}) > 0$, thus the process is always dissipative. Hence, we have accomplished the proof that particles moving in the Galileo-Newton vacuum medium which yields the particle energy spectra without energy gaps will feel dissipative force and the Galileo-Newton vacuum medium is a dissipative medium rather than a super fluid medium.

On the other hand, as the particle is rest with respect to the Galileo-Newton vacuum medium,. it will remain rest for ever because of the energy-momentum conservation. Thus the moving particle and the rest particle have different physical behaviors and the Galileo-Newton's principle of relativity breaks down for the Galileo-Newton vacuum. To make the Galileo-Newton's principle of relativity valid, the Galileo-Newton vacuum must be absolutely vacant so that no medium can leads to dissipation. That is why Newton made the assumption that the Galileo-Newton vacuum (space ) is absolutely empty for the consistency of his theory of classical mechanics.

In conclusion, to observe the principle of relativity and the energy-momentum conservation law, the Galileo-Newton vacuum must be assumed to be absolutely empty and the Lorentz-Einstein vacuum must be full of super fluid medium. Since within the Poincare group, the translation invariance leads to the energy-momentum conservation and the Lorentz invariance leads to the relativistic energy-momentum dispersion relation, the super fluid property of the Lorentz-Einstein vacuum medium is an inevitable consequence of the Poincare invariance of the vacuum medium.

**Acknowledgements**. This work was supported in part by the National Natural Science Foundation of China under the grant No. 10375039 and 90503008, the Doctoral Education Fund of the Education Ministry of China, and by the Research Fund of the Nuclear Theory Center of HIRFL of China.

**references**
1. Landau,L.D.&Lifshits, E.M., *Statistical Physics,* Chapter 6, (Pergamon Press, 1958.)
2. Landau,L.D.&Lifshits, E.M., *Statistical Physics Part 2,* Chapter 3, *third edition*, (Betterworth-Heineman, 1999).